# Generalized quantification of three-dimensional resolution in optical diffraction tomography using the projection of maximal spatial bandwidths

CHANSUK PARK,[1,2] SEUNGWOO SHIN,[1,2] AND YONGKEUN PARK[1,2,3,*]

[1]Department of Physics, Korea Advanced Institutes of Science and Technology (KAIST), Daejeon 34141, Republic of Korea
[2]KAIST Institute for Health Science and Technology, KAIST, Daejeon 34141, Republic of Korea
[3]Tomocube, Inc., Daejeon 34051, Republic of Korea
*Corresponding author: yk.park@kaist.ac.kr

**Optical diffraction tomography (ODT) is a three-dimensional (3D) quantitative phase imaging technique, which enables the reconstruction of the 3D refractive index (RI) distribution of a transparent sample. Due to its fast, non-invasive, and quantitative 3D imaging capability, ODT has emerged as one of the most powerful tools for various applications. However, the spatial resolution of ODT has only been quantified along the lateral and axial directions for limited conditions; it has not been investigated for arbitrary-oblique directions. In this paper, we systematically quantify the 3D spatial resolution of ODT by exploiting the spatial bandwidth of the reconstructed scattering potential. The 3D spatial resolution is calculated for various types of ODT systems, including the illumination-scanning, sample-rotation, and hybrid scanning-rotation methods. In particular, using the calculated 3D spatial resolution, we provide the spatial resolution as well as the arbitrary sliced angle. Furthermore, to validate the present method, the point spread function of an ODT system is experimentally obtained using the deconvolution of a 3D RI distribution of a microsphere and is compared with the calculated resolution. The present method for defining spatial resolution in ODT is directly applicable for various 3D microscopic techniques, providing a solid criterion for the accessible finest 3D structures.**

## 1. INTRODUCTION

Optical resolution is a measure of the ability to resolve fine details in an image, and is fundamentally limited by diffraction limit or the spatial bandwidth of an imaging system. Due to the limited spatial resolution, a point source is imaged as an enlarged spot, called the point spread function (PSF). In a linear shift-invariant imaging system, an outcome image is expressed as the convolution of an input image with the PSF of an imaging system.

In the 19th century, Rayleigh defined the spatial resolution as a minimum resolvable distance between two PSFs from two point sources [1]. This criterion is valid for an incoherent imaging system because two incoherent PSFs do not interfere with each other. However, for a coherent imaging system, two coherent PSFs interfere, and the resultant image depends on both the distance between the two sources and relative phase between PSFs [2]. Meanwhile, Abbe established the diffraction limit using the spatial bandwidth of an imaging system [3]. The resolution in the Abbe limit can be quantified as $\lambda/2/NA$, where $\lambda$ is the wavelength of illumination and $NA$ is the numerical aperture (NA) of an imaging system. The Abbe limit is the inverse of the spatial bandwidth of an optical transfer function (OTF).

These precedent criteria for quantifying spatial resolutions are also applicable to three-dimensional (3D) microscopic techniques, including confocal microscopy, optical coherence tomography, X-ray computed tomography (CT), electron crystallography, and optical diffraction tomography (ODT). Although various studies have addressed the spatial resolution of 3D microscopy, the spatial resolution of 3D microscopic techniques has not yet been fully investigated.

Previously, the resolutions of various 3D microscopes were quantified by experimentally measuring PSFs [4-7], and the corresponding OTFs of 3D microscopes are well defined by the analytical forms [8-14]. In most cases, only the lateral and axial resolutions have been discussed for specific imaging systems, while the resolution along an arbitrary angle remains unexplored.

Here, we systematically investigate the spatial resolutions along arbitrary directions exploiting projected spatial bandwidths in 3D Fourier space. We propose a general approach to quantify the spatial resolution in an arbitrary-angled plane by exploiting the Fourier slice theorem and Abbe limit. Using the proposed method, we quantify the resolution of ODT systems across various methods and experimental environments. Although we used ODT to demonstrate the applicability, the present method is sufficiently general and can be readily applied to other various 3D microscopy techniques.

## 2. METHODS

### A. Optical diffraction tomography

ODT is a 3D quantitative phase imaging (QPI) technique, which measures the 3D refractive index (RI) distribution of a transparent microscopic sample, such as biological cells [15-17]. ODT is an optical analogy to X-ray CT; the 3D tomogram of a sample is reconstructed from multiple measurements of 2D images via the inverse scattering principle.

Generally, optical systems for ODT consists of a (i) QPI part to measure the 2D optical fields of a sample and (ii) device to control the angle of a coherent illumination impinging onto a sample. The control of the illumination angle has been achieved using a galvanometric mirror [8, 18], spatial light modulator [19], or digital micromirror device [20, 21]. Alternatively, the rotation of a sample [22, 23] or the wavelength scanning of an illumination beam [24] can also be used to achieve ODT. Recently, white-light illumination in combination with axial scanning was also used for ODT [25].

ODT provides label-free and quantitative imaging of live cells and tissues without complicated sample preparation procedures [26]. ODT has been widely utilized for the study of biological samples, including microalgae [27], plant biology [28], membrane biophysics [29], immunology[30], infectious disease [31], and pharmacology [32].

### B. Spatial resolution in an arbitrary plane

Because ODT provides the morphology of various subcellular organelles in 3D, it has become increasingly important for examining micrometer-sized objects, not only in a lateral direction, but also in the arbitrary direction. As shown in Fig. 1(a), the ODT image of a hepatocyte show various subcellular structures including nucleus membrane, nucleoli, and lipid droplets [33]. However, the resolution of ODT has been studies mostly in the lateral or axial directions in a specific limited condition of optical imaging.

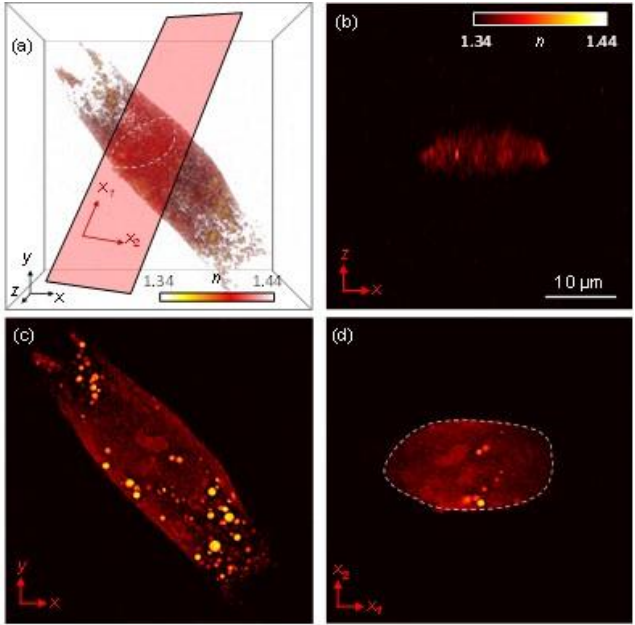

Fig. 1. (a) 3D rendered isosurface of the 3D RI distribution of a hepatocyte. (b and c) Cross-sectional slices of the 3D RI distribution of the cell in the *x-z* and *x-y* planes. (d) Cross-sectional slice of the 3D RI distribution of the cell in an oblique plane as represented by the $x_1$ and $x_2$ arrows in (a).

Due to the anisotropicity of OTF in 3D optical microscopes including ODT, the spatial resolution of 3D microscopes needs to be defined in arbitrary planes. This is because the collectible spatial bandwidth in 3D microscopes has a directional dependence [11, 34, 35]. In particular, most 3D microscopes suffer from a poor frequency cover range in the axial direction. Likewise, the resolution in ODT is restrictively quantified only for the lateral and axial directions. Conventional studies on ODT present an OTF of an ODT system with restrictively quantified spatial bandwidths along the lateral and axial directions [Figs. 1(b) – 1(c)] [8, 36]. However, the resolution in the oblique plane as in Fig. 1(d) remains unclear.

## C. Different methods for ODT and their OTFs

Using measured multiple 2D optical fields, ODT reconstructs a 3D scattering potential in Fourier or spatial frequency space. According to Fourier diffraction theorem, each 2D optical field is mapped onto an Ewald sphere in Fourier space [15]. For each illumination angle, only a fraction of the Ewald sphere is measured and mapped due to the limited NA of the objective lens. By sequentially mapping each Ewald cap according to a corresponding illumination angle, the 3D scattering potential is reconstructed from which the 3D RI tomogram of a sample is retrieved via an inverse Fourier transformation.

Various ODT approaches have been proposed that include the measuring of multiple 2D optical fields. Among them, three representative approaches; the illumination scanning, sample rotation, and hybrid scanning-rotation methods, will be discussed in this work (Fig. 2).

In the illumination scanning method [Fig. 2(a)], the scattered fields from a sample are measured with various illumination angles (represented by its spatial frequency, $\vec{v}_i$). Each Ewald cap is mapped So that its center is positioned at $-\vec{v}_i$ [Figs. 2(b)–2(c)] [15].

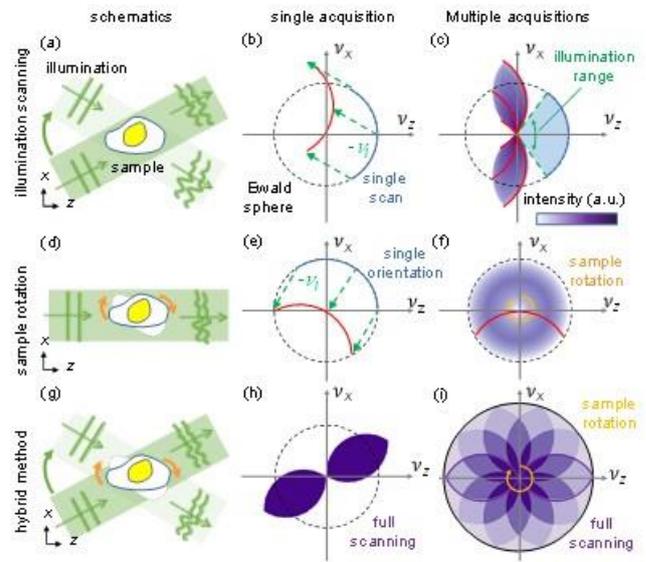

Fig. 2. Schematic illustration of various ODT methods and corresponding Fourier spectra. (a–c) Illumination scanning method. (a) Fields scattered from various illumination angles are recorded. (b) Reconstruction procedure with a single Ewald cap. (c) Fourier spectra of an illumination scanning method. (d–f) Sample rotation method. (d) A sample is rotated while the illumination angle is fixed. (e) Reconstruction procedure with a single Ewald cap. (f) Fourier spectra of a sample rotation method. (g–i) Hybrid scanning-rotation method. (g) Both the illumination angle and sample orientation are controlled. (h) Fourier spectra synthesized from multiple illumination angles at a specific sample orientation. after the reconstruction. (i) Fourier spectra of a hybrid scanning-rotation method.

Depending on the sampling of $\vec{v}_i$, the illumination scanning method can be further classified. In this work, we address (i) circular scanning and (ii) full mesh scanning. The full mesh scanning utilizes all available $\vec{v}_i$ within the NA of a condenser lens, whereas the circular scanning utilizes $\vec{v}_i$ along the NA circle edge. Illumination scanning is the most commonly used technique in ODT because it is relatively easy to implement and can minimize the alteration of a sample. However, its image quality is degraded due to the inaccessible information along the axial direction, which is also known as the missing cone [34, 37].

In the sample rotation method [Fig. 2(d)], the orientation of a sample is rotated with respect to the axis parallel to the focal plane, while the illumination beam is fixed normal to the focal plane [22, 38, 39]. As the sample is rotated, multiple 2D optical fields are measured, which are then mapped onto Ewald caps [Figs. 2(e)–(f)]. The 3D OTF of the sample rotation method is almost spherical, but also has missing frequency information along the rotation axis, which is also known as the 'missing apple core' [40]. The weakness of the sample rotation method is its difficult implementation because the stable and accurate rotation of a live biological cell is experimentally challenging.

The hybrid scanning-rotation method combines illumination scanning with sample rotation by conducting illumination scanning in different sample orientations [Fig. 2(g)] [36]. If illumination scanning is thoroughly conducted for all possible sample orientations, the scattering potential of a sample in frequency space is obtainable in a full spherical shape [Figs. 2(h) and 2(i)]. Thus, the 3D OTF of the hybrid method also becomes a perfect sphere with an isotropic spatial bandwidth, which is larger than that of the mesh scanning method.

The lateral and vertical spatial bandwidths of the OTFs from various methods expressed in analytical form [8,36] are given below:

$$\Delta v_{x,y}^{Mesh} = 2n_m v_0 \left(\sin\theta_i + \sin\theta_o\right), \Delta v_z^{Mesh} = n_m v_0 \left(2 - \cos\theta_i - \cos\theta_o\right),$$
$$\Delta v_{x,y}^{Circle} = 2n_m v_0 \left(\sin\theta_i + \sin\theta_o\right), \Delta v_z^{Circle} = n_m v_0 \left(1 - \cos\theta_o\right),$$
$$\Delta v_{x,z}^{Sample} = 4n_m v_0 \sin\left(\theta_o/2\right), \Delta v_y^{Sample} = 2n_m v_0 \sin\theta_o,$$
$$\Delta v_{x,y,z}^{Hybrid} = 4n_m v_0 \sin\left[\left(\theta_i + \theta_o\right)/2\right],$$
(1)

where nm is the refractive index of a medium; $v_0$ is the spatial frequency of an illumination beam in vacuum; and $\theta_i$ and $\theta_o$ are angles obtained from the NA of condenser ($NA_i$) and objective ($NA_o$) lenses, respectively, using the relation $NA = n \sin\theta$. For example, mesh scanning with $NA_i = NA_o = 1.25$, $n = 1.52$, and $\lambda = 6.3$ nm gives the lateral resolution of 126 nm and axial resolution of 483 nm using Eq. (1) [8].

### D. Projected spatial bandwidth method

The present method for quantifying the resolution in an arbitrary plane is an expansion of the Abbe limit to the 3D case. The spatial resolution is quantified along a specific direction by taking the inverse of the maximum achievable spatial bandwidth.

For a given ODT system with the 3D OTF $O(\vec{v})$, a reconstructed 3D scattering potential S is given by $S(\vec{r}) = k_0^2 \{n^2(\vec{r}) - n_m^2\}$ [15]. For a transparent biological sample, its RI can be assumed real. Then, the conjugate symmetry of the sample satisfies,

$$\tilde{S}(\vec{v}) = \tilde{S}^*(-\vec{v}),$$
(2)

where $\tilde{S}$ and $*$ denote the 3D Fourier transformation of $S$ and the complex conjugation, respectively. Using Eq. (2), the following equation for the reconstructed scattering potential of a sample, $\text{Re}[S_{obs}(\vec{r})]$, is (see Appendix A),

$$\text{Re}[S_{obs}(\vec{r})] = \text{Re}\left[\int O(\vec{v})\tilde{S}(\vec{v})\exp(j\vec{v}\cdot\vec{x})dv^3\right]$$
$$= \text{Re}\left[\frac{1}{2}\int\{O(\vec{v}) + O(-\vec{v})\}\tilde{S}(\vec{v})\exp(j\vec{v}\cdot\vec{x})dv^3\right].$$
(3)

Then, the effective OTF for a transparent sample, $O_{eff}$, is given by

$$O_{eff}(\vec{v}) = \frac{1}{2}[O(\vec{v}) + O(-\vec{v})].$$
(4)

Once $O_{eff}$ is obtained, the spatial bandwidth $\Delta v_1$ along an arbitrary direction in 3D can be retrieved. According to Fourier slice theorem, a cross-sectional image in 3D image space is equal to the Fourier transform of the projection of the sample information onto the very plane in Fourier space [Figs. 3(b) and 3(c)]. Thus, from the projected OTF of an imaging system onto an arbitrary direction $\vec{v}_1$, the maximum achievable spatial bandwidth along the direction $\Delta v_1$ can be obtained. Using the obtained frequency, we can define the spatial resolution in the arbitrary direction $d_1$, as Abbe did in the 2D case: $d_1 = 1/\Delta v_1$.

Repeating the projection procedure over different angles, a set of resolutions of the corresponding ODT system in various angles can be retrieved, from which the 3D resolution of the ODT system is quantified [Fig. 3(d)]. The 3D resolution is presented as a function of the polar, $\theta$, and azimuthal, $\phi$, angles of $\vec{v}_1$.

Numerical experiments were conducted to obtain the 3D resolution of various ODT methods. To effectively consider various combinations of the illumination and detection NAs, the 3D resolution quantification for each ODT method was simulated with three different partial coherence parameters $\sigma = NA_i / NA_o = $ (100%, 90%, and 70%) [41].

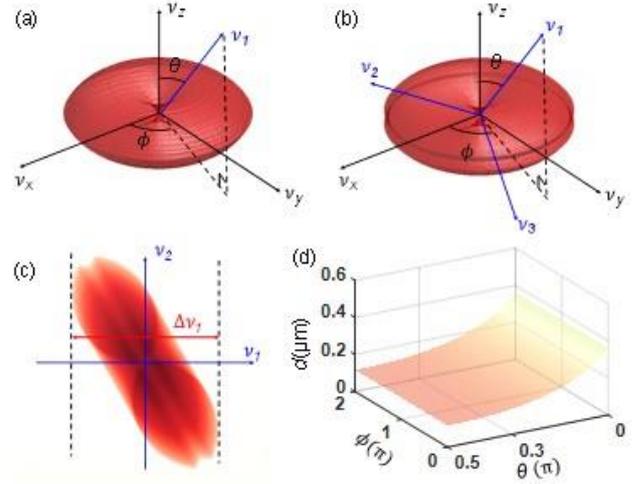

Fig. 3. (a) 3D OTF of an ODT system. To define the spatial resolution along the $x_1$ direction, the spatial bandwidth of the 3D OTF along the $v_1$ direction is obtained. The polar and azimuthal angle are denoted by $\theta$ and $\phi$, respectively. (b) The spatial bandwidth in an arbitrary direction $\vec{v}_1$ is obtained from the projection of OTF onto $\vec{v}_1$. (c) Projection of 3D OTF onto the $v_1$–$v_2$ plane. (d) Calculated bandwidths along all $\theta$ and $\phi$.

## 3. RESULTS

As discussed in Section 2.B, the 3D resolution of ODT is systematically quantified and analyzed through numerical simulations using various OTFs and $\sigma$. For various ODT systems, the 3D resolution is quantified as a function of $\theta$ and $\phi$. The following values were used for the simulation: $\lambda = 532$ nm, $n_m = 1.337$, and $NA_o = 1.2$.

### A. 3D resolution in ODT

The 3D resolution of the ODT systems for various methods (columns) and coherence parameters (rows) are shown in Fig. 4.

The illumination scanning methods (mesh and circular scanning) have minimum and maximum resolutions of 0.11 and 0.59 μm, respectively [Figs. 4(a)–4(f)]. The minimum resolution is achieved for the lateral direction ($\theta = 0.5\pi$), and the resolution gradually increases towards the axial direction ($\theta = 0$). The poor resolution at $\theta = 0$ is a consequence of the short spatial bandwidth of the OTF along the axial direction. The intriguing point in the illumination scanning is that the 3D resolution of the mesh scanning and circular scanning coincides at $\sigma = 100\%$. When the NAs of the condenser and objective lens are the same, the resolution is 0.11 μm at $\theta = 0$ and 0.35 μm at $\theta = 0.5\pi$ for both scanning methods [Figs. 4(a) and 4(d)]. The same 3D resolutions are obtained from the same effective OTF of the two ODT systems at $\sigma = 100\%$. However, the resolving ability of the circular scanning method degrades rapidly as $\sigma$ decreases, and the effective OTFs from mesh scanning and circular scanning start to differ. The circular scanning method has an axial resolution (where the resolution is the poorest) of 0.35, 0.48, and 0.59 μm at $\sigma = 100, 90$, and $70\%$, respectively [Figs. 4(d)–4(f)]. On the contrary, there is no recognizable regression of resolution in the mesh scanning method as σ changes (only by a few nm) [Figs. 4(a)–4(c)].

The sample rotation method presents a uniform 3D resolution of 0.19 μm, except for the small angular region near the rotation axis (the y-axis) where $(\theta, \phi) = (0.5\pi, 0.5\pi)$ or $(0.5\pi, 1.5\pi)$ [Figs. 4(g)–4(i)]. The resolution is 0.22 μm at the sample rotation axis. σ does not affect the resolution of the sample rotation method since the angle of the illumination beam is fixed normal to the sample plane.

The spatial resolution of the sample rotation method is smaller than the illumination scanning method in most directions, but the sample rotation has a larger minimum resolution compared with the illumination scanning method.

The hybrid scanning-rotation method presents uniform 3D resolutions of 0.11, 0.12, and 0.13 μm at σ = 100, 90, and 70% respectively [Figs. 4(j)–4(l)]. The resolution of the hybrid method is less than that of the minimum resolution of the illumination scanning, as expressed by Eq. (5);

$$\Delta v_{x,y}^{\text{Mesh}} \left(= 2n_m v_0 (\sin\theta_i + \sin\theta_o)\right) \leq \Delta v_{x,y,z}^{\text{Hybrid}} \left(= 4n_m v_0 \sin[(\theta_i + \theta_o)/2]\right). \quad (5)$$

Similar to the sample rotation method, the hybrid scanning-rotation method is difficult to implement, however, an enhancement in the spatial resolution can be achieved [36]

### B. Experimental verification

To demonstrate the feasibility of the simulated OTFs in an experimental situation, we compared theoretical PSFs and the experimentally measured PSFs of an ODT system.

The experimental PSF, $P_{\text{exp}}(\vec{r})$, was obtained by performing the deconvolution between the reconstructed scattered field and *priori* sample information, as described in Eq. (6) [42];

$$P_{\text{exp}}(\vec{r}) = \tilde{O}_{\text{exp}}(\vec{r}) = iFT\left[\tilde{S}_{obs}(\vec{v}) / \tilde{S}(\vec{v})\right], \quad (6)$$

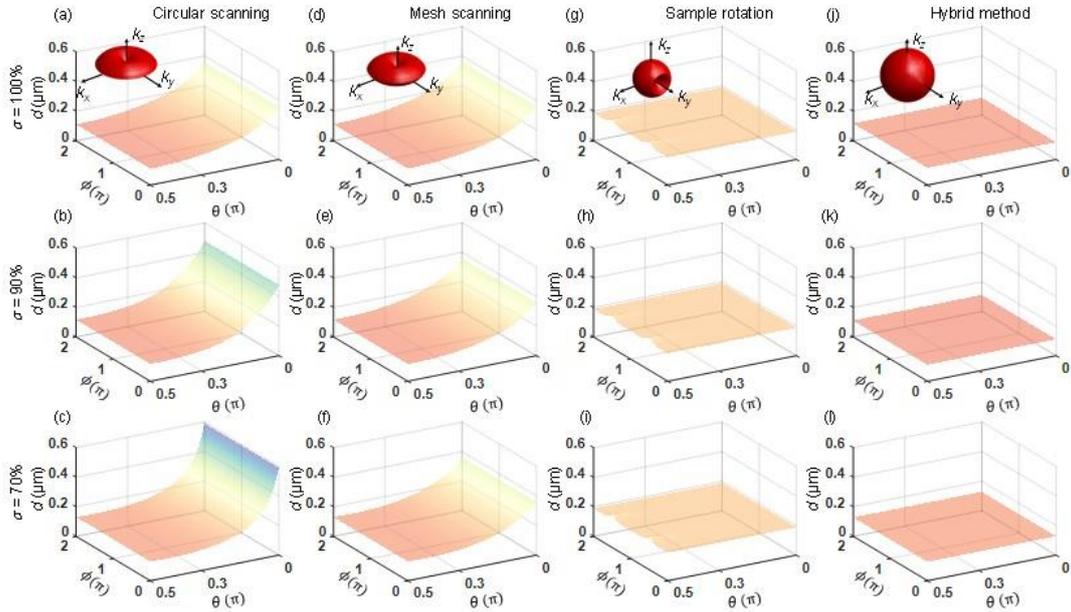

Fig. 4. 3D resolutions of ODT as a function of $\theta$ and $\phi$, obtained for various beam illuminations and partial coherence parameters. (a–c) Mesh scanning. (d–f) Circular scanning. (g–i) Sample rotation. (j–l) Hybrid method. The OTFs of various ODT methods are shown as insets in the first row.

where iFT denotes the inverse Fourier transformation and $\tilde{S}_{obs}(\vec{v})$ is an experimentally measured scattering potential corresponding to the sample information given by Eq. (1).

The result is shown in Fig. 5. A silica bead was experimentally measured [Fig. 5(a)], and the 3D resolution of the used ODT system was investigated using the measured RI tomogram. Detailed information relating to the experimental is described in Appendix B and Ref. [37]. A phantom of the bead $\tilde{S}(\vec{v})$ was generated using information regarding the microsphere provided by the manufacturer (n = 1.34 and a diameter of 5 μm) [Fig. 5(b)].

Before the deconvolution process, $\tilde{S}(\vec{v})$ was band-pass filtered in frequency space to avoid divergence during the division in Eq. (6): the absolute value of $\tilde{S}(\vec{v})$ was set to one, and its absolute value was smaller than $10^{-2.65}$ of the maximum $|\tilde{S}(\vec{v})|$. The phase of $\tilde{S}(\vec{v})$ was conserved during the filtering process.

The threshold value was determined so that 89% of the original bandwidth was utilized after the band-pass filtering. Conversely, the theoretical PSF, $P_{\text{theory}}(\vec{r})$, was obtained by taking the inverse Fourier transformation of the theoretical OTF, $O_{\text{theory}}(\vec{v})$;

$$P_{\text{theory}}(\vec{r}) = iFT\left[O_{\text{theory}}(\vec{v})\right]. \quad (7)$$

The experimental PSF, together with the theoretical PSF is shown for various polar angles in Fig. 5(d). The experimentally measured resolutions in 3D are consistent with those expected from the numerical simulation. For most of the angles, the two central peaks of pairs of PSFs overlap, with consistent widths. The discrepancy between the two PSFs are most significant in the axial direction ($\theta$ = 0), where the optical aberration is the most significant. Another observable difference is the side lobes in the experimental PSFs. The diverging voxels create side lobes in frequency space, which was not eliminated in the band-pass filtering process. Other errors may come from the sparsely measured high-frequency region, where noises are not completely averaged out.

## 4. DISCUSSIONS

We present a general approach to quantify the 3D resolution of ODT. Exploiting 3D OTFs and the maximum achievable spatial frequency, the resolutions along arbitrary directions are systematically investigated for various types of ODT. The present method is verified by an experiment using a silica microsphere. Our method provides a quantitative and intuitive description of the 3D imaging capabilities of microscopes, which was not achieved by simple presentation of the 3D PSF or OTF.

The present technique is sufficiently broad and general, offering approaches for the prediction and quantification of other 3D QPI techniques, including 3D ptychography and white-light diffraction tomography [25, 43, 44]. It serves as an important method in diverse applications. Furthermore, because the present method is based on 3D OTF, it can be applied, not only to coherent microscopy, but also to various types of incoherent 3D microscopy [12-14].

Although this work focuses on a transparent object, the approach can also be expanded to absorptive samples. To obtain the absorption tomogram of a sample, the imaginary parts can be retrieved from the reconstructed 3D tomogram of complex RI values. For absorptive samples, the resolution can be quantified using the original OTF of an ODT system. To apply the present method to absorptive samples, aberrations in an ODT system should be minimized so that the OTF is point symmetric about the origin. Otherwise, the corresponding PSF becomes a complex-valued function, resulting in an inaccurate quantification of 3D resolution.

The present method is also applicable to partially coherent ODT (PC-ODT) or incoherent ODT. In PC-ODT, partially coherent light impinges onto a sample, and the 2D images from various $z$-planes are collected to reconstruct the 3D RI tomogram [45-47]. The band-pass filtering process is performed before the reconstruction of the scattering potential of a sample, so that the OTF of the PC-ODT is discrete, unlike the OTFs in coherent ODT. Because the proposed method defines the resolution from the cut-off spatial bandwidth, the artifacts from discrete OTF may not be included.

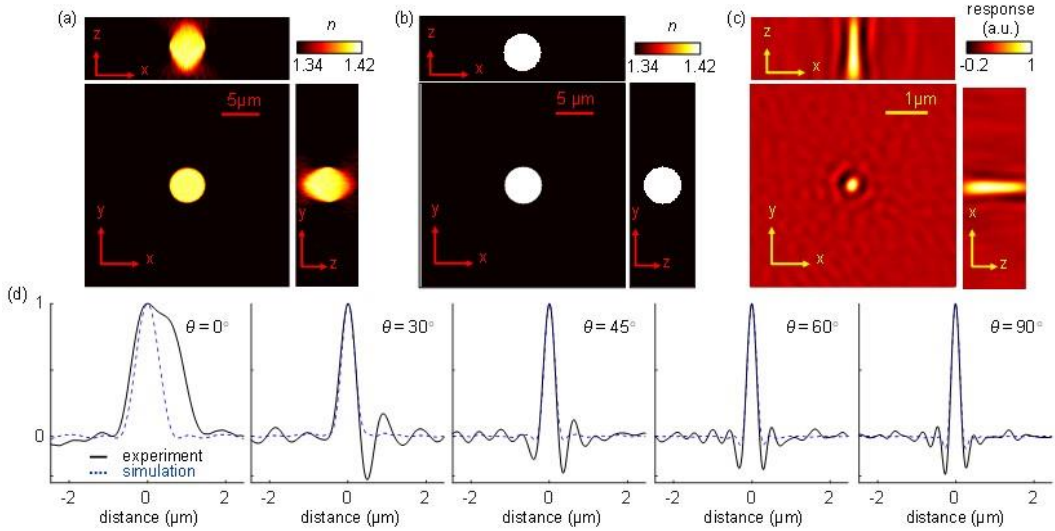

Fig. 4. Experimental assessment. (a) Cross-sectional slices of the reconstructed RI map of a silica bead in the $x$–$y$, $x$–$z$, and $y$–$z$ planes. (b) Phantom of an ideal 5-μm-diameter silica bead. (c) Experimental PSF of the optical system acquired using the deconvolution of the experimentally measured tomogram and the phantom. (d) Experimental and theoretical PSF viewed at various polar angles.

It should be noted that the regularization or discretization methods [34, 37, 48] that have been used to enhance the image quality in ODT by filling the missing cone information, can alter the maximum spatial bandwidth. Thus, the use of this regularization should be carefully performed, particularly when addressing the 3D resolution of ODT systems using the present method.

Nonetheless, the present method has several limitations. Because it is based on the projection of the maximum spatial bandwidth, not on the transfer values of spatial frequencies, it does not consider the effects of the inhomogeneity of an OTF projection. Thus, our method assumes the ideal OTF of an imaging system; it does not address the effects from the missing cones or other various artifacts from OTF geometries as in the discrete OTF of PC-ODT. Morover, as for other imaging techniques, the present method is based on the assumption of the linear shift invariant – which is a prerequisite for the definition of PSF or OTF. Severe aberration may result in the violation of this assumption. Nonetheless, the present method can also be extended to such a case when combined with a transmission matrix approach, providing information regarding the position-dependent PSF or OTF [49].

## 5. CONCLUSIONS

In conclusion, we have established a full-angle resolution quantification method for an ODT system. The resolution of the ODT system for an arbitrary, oblique angle was quantified by taking an inverse of the spatial bandwidth in the corresponding direction. The spatial bandwidth in the direction of interest was obtained by projecting the effective OTF at the same direction. The set of resolution values obtained at various angles was defined as the 3D resolution of the ODT system.

Using the present method, the 3D resolutions of ODT systems with different methods and coherence parameters were quantified. The results provide the resolution of ODT in all 3D directions, which was not explicitly measured in the past.

Furthermore, the validation of the theoretical OTF used for the simulation was conducted by comparing experimental and theoretical PSFs. Despite the presence of noises and aberration arising from the unideal imaging system, the central peak of the PSFs overlapped in the central region, proving the validity of the ideal OTF used in the simulation.

Although the demonstration of the present method was limited to ODT systems, the method also applies to various 3D microscopes using their OTFs. We expect that this work will provide solid criterion in the resolution of 3D microscopes for various studies.

## APPENDIX A: Effective OTF

In ODT, the obtained reconstructed scattering potential is expressed as a product of an original scattering potential, $\tilde{S}(\vec{v})$, and the OTF of an imaging system, $O(\vec{v})$, as follows;

$$\tilde{S}_{obs}(\vec{v}) = O(\vec{v})\tilde{S}(\vec{v}). \quad (A1)$$

The real part of the sample information is then obtained by taking the inverse Fourier transformation.

$$\text{Re}\left[S_{obs}(\vec{r})\right] = \text{Re}\left[\int O(\vec{v})\tilde{S}(\vec{v})\exp(i\vec{r}\cdot\vec{v})dv^3\right]. \quad (A2)$$

The real and imaginary parts of the sample scattering potential can be written separately as $\tilde{S}(\vec{v}) = \tilde{S}_r(\vec{v}) + j\tilde{S}_i(\vec{v})$, and by the conjugate symmetric property of the real-valued sample $\tilde{S}(-\vec{v}) = \tilde{S}^*(\vec{v})$, we get $\tilde{S}_r(\vec{v}) = \tilde{S}_r(-\vec{v})$ and $\tilde{S}_i(\vec{v}) = -\tilde{S}_i(-\vec{v})$. Equation (A2) can then be rewritten as

$$\begin{aligned}\text{Re}\left[S_{obs}(\vec{r})\right] &= \int O(\vec{v})\left[\tilde{S}_r(\vec{v})\cos(\vec{v}\cdot\vec{x}) - \tilde{S}_i(\vec{v})\sin(\vec{v}\cdot\vec{x})\right]dv^3 \\ &= \int \tfrac{1}{2}\left[O(\vec{v})+O(-\vec{v})\right]\left[\tilde{S}_r(\vec{v})\cos(\vec{v}\cdot\vec{x}) - \tilde{S}_i(\vec{v})\sin(\vec{v}\cdot\vec{x})\right]dv^3 \\ &= \int O_{eff}(\vec{v})[\tilde{S}_r(\vec{v})\cos(\vec{v}\cdot\vec{x}) - \tilde{S}_i(\vec{v})\sin(\vec{v}\cdot\vec{x})]dv^3 \\ &= \text{Re}\left[\int O_{eff}(\vec{v})\exp(j\vec{v}\cdot\vec{x})dv^3\right], \end{aligned} \quad (A3)$$

where the effective OTF is defined as $O_{eff}(\vec{v}) = \tfrac{1}{2}\left[O(\vec{v})+O(-\vec{v})\right]$.

## APPENDIX B: Experimental setup

The schematic of the used ODT setup used is shown in Fig. A1. The system is based on off-axis Mach-Zehnder interferometry equipped with a DMD.

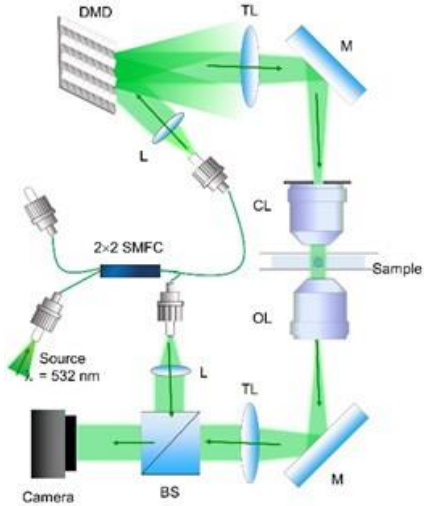

Fig. A1. Schematic of the optical setup. M: mirror, L: convex lens, CL: condenser lens, OL: objective lens, TL: tube lens, BS: beam splitter, and SMFC: single mode fiber coupler.

The plane wave from a diode-pumped solid-state laser ($\lambda$ = 532 nm, 50 mW, Cobolt Co., Solna, Sweden) was split into the sample and reference arm using a fiber coupler (TW560R2F2, Thorlabs, Inc., USA). The DMD (DLP LightCrafter 3000, Texas Instruments, Inc., USA) was used to control the angle of the illumination beam impinging onto a sample [20]. The beam diffracted from the DMD was illuminated onto a sample using a condenser lens (UPLASAPO 60XW, Olympus Inc., Japan). Then, the scattered beam from a sample was collected using an objective lens (UPLASAPO 60XO, Olympus Inc., Japan). The NAs of the objective and condenser lenses were 1.42 and 0.9, respectively. Then, both the sample and reference beams were combined by a beam splitter and projected onto a CCD camera (FL3-U3-13Y3M-C, FLIR Systems, Inc., USA), where spatially modulated holograms were recorded. In total, 50 2-D optical fields were used to reconstruct a 3D RI tomogram.

## Funding
This work was supported by KAIST, BK21+ program, Tomocube, and National Research Foundation of Korea (2015R1A3A2066550, 2017M3C1A3013923, 2014K1A3A1A09063027).

## Acknowledgment
We thank Dr. Kyoohyun Kim for helpful discussions in Section 2.

## References

1. Rayleigh, "XXXI. Investigations in optics, with special reference to the spectroscope," The London, Edinburgh, and Dublin Philosophical Magazine and Journal of Science **8**, 261-274 (1879).
2. R. Horstmeyer, R. Heintzmann, G. Popescu, L. Waller, and C. Yang, "Standardizing the resolution claims for coherent microscopy," Nature Photonics **10**, 68 (2016).
3. E. Abbe, "Beiträge zur Theorie des Mikroskops und der mikroskopischen Wahrnehmung," Archiv für mikroskopische Anatomie **9**, 413.
4. S. R. P. Pavani, M. A. Thompson, J. S. Biteen, S. J. Lord, N. Liu, R. J. Twieg, R. Piestun, and W. E. Moerner, "Three-dimensional, single-molecule fluorescence imaging beyond the diffraction limit by using a double-helix point spread function," Proceedings of the National Academy of Sciences **106**, 2995-2999 (2009).
5. R. W. Cole, T. Jinadasa, and C. M. Brown, "Measuring and interpreting point spread functions to determine confocal microscope resolution and ensure quality control," Nature Protocols **6**, 1929 (2011).
6. H. KIRSHNER, F. AGUET, D. SAGE, and M. UNSER, "3-D PSF fitting for fluorescence microscopy: implementation and localization application," Journal of Microscopy **249**, 13-25 (2013).
7. T. A. Klar, S. Jakobs, M. Dyba, A. Egner, and S. W. Hell, "Fluorescence microscopy with diffraction resolution barrier broken by stimulated emission," Proceedings of the National Academy of Sciences **97**, 8206-8210 (2000).
8. V. Lauer, "New approach to optical diffraction tomography yielding a vector equation of diffraction tomography and a novel tomographic microscope," Journal of Microscopy **205**, 165-176 (2002).
9. M. G. L. Gustafsson, D. A. Agard, and J. W. Sedat, "Sevenfold improvement of axial resolution in 3D wide-field microscopy using two objective lenses," in *IS&T/SPIE's Symposium on Electronic Imaging: Science and Technology*, (SPIE, 1995), 10.
10. M. R. Arnison and C. J. R. Sheppard, "A 3D vectorial optical transfer function suitable for arbitrary pupil functions," Optics Communications **211**, 53-63 (2002).
11. O. Nakamura and S. Kawata, "Three-dimensional transfer-function analysis of the tomographic capability of a confocal fluorescence microscope," J. Opt. Soc. Am. A **7**, 522-526 (1990).
12. S. Kimura and C. Munakata, "Calculation of three-dimensional optical transfer function for a confocal scanning fluorescent microscope," J. Opt. Soc. Am. A **6**, 1015-1019 (1989).
13. C. J. R. Sheppard and M. Gu, "The significance of 3-D transfer functions in confocal scanning microscopy," Journal of Microscopy **165**, 377-390 (1992).

14. S. Lindek and E. H. K. Stelzer, "Optical transfer functions for confocal theta fluorescence microscopy," J. Opt. Soc. Am. A **13**, 479-482 (1996).
15. E. Wolf, "Three-dimensional structure determination of semi-transparent objects from holographic data," Optics Communications **1**, 153-156 (1969).
16. G. Popescu, *Quantitative phase imaging of cells and tissues* (McGraw Hill Professional, 2011).
17. K. Kim, J. Yoon, S. Shin, S. Lee, S.-A. Yang, and Y. Park, "Optical diffraction tomography techniques for the study of cell pathophysiology," Journal of Biomedical Photonics & Engineering **2**, 020201 (2016).
18. Y. Sung, W. Choi, C. Fang-Yen, K. Badizadegan, R. R. Dasari, and M. S. Feld, "Optical diffraction tomography for high resolution live cell imaging," Optics express **17**, 266-277 (2009).
19. A. Kuś, W. Krauze, and M. Kujawińska, "Active limited-angle tomographic phase microscope," Journal of biomedical optics **20**, 111216 (2015).
20. S. Shin, K. Kim, J. Yoon, and Y. Park, "Active illumination using a digital micromirror device for quantitative phase imaging," Optics Letters **40**, 5407-5410 (2015).
21. K. Kim, Z. Yaqoob, K. Lee, J. W. Kang, Y. Choi, P. Hosseini, P. T. C. So, and Y. Park, "Diffraction optical tomography using a quantitative phase imaging unit," Opt Lett **39**, 6935-6938 (2014).
22. F. Charrière, A. Marian, F. Montfort, J. Kuehn, T. Colomb, E. Cuche, P. Marquet, and C. Depeursinge, "Cell refractive index tomography by digital holographic microscopy," Optics letters **31**, 178-180 (2006).
23. M. Habaza, B. Gilboa, Y. Roichman, and N. T. Shaked, "Tomographic phase microscopy with 180° rotation of live cells in suspension by holographic optical tweezers," Optics letters **40**, 1881-1884 (2015).
24. J. Kühn, F. Montfort, T. Colomb, B. Rappaz, C. Moratal, N. Pavillon, P. Marquet, and C. Depeursinge, "Submicrometer tomography of cells by multiple-wavelength digital holographic microscopy in reflection," Optics letters **34**, 653-655 (2009).
25. T. Kim, R. Zhou, M. Mir, S. D. Babacan, P. S. Carney, L. L. Goddard, and G. Popescu, "White-light diffraction tomography of unlabelled live cells," Nature Photonics **8**, 256 (2014).
26. D. Kim, S. Lee, M. Lee, J. Oh, S.-A. Yang, and Y. Park, "Refractive index as an intrinsic imaging contrast for 3-D label-free live cell imaging," bioRxiv, 106328 (2017).
27. J. Jung, S.-J. Hong, H.-B. Kim, G. Kim, M. Lee, S. Shin, S. Lee, D.-J. Kim, C.-G. Lee, and Y. Park, "Label-free non-invasive quantitative measurement of lipid contents in individual microalgal cells using refractive index tomography," Scientific Reports **8**, 6524 (2018).
28. G. Kim, S. Lee, S. Shin, and Y. Park, "Three-dimensional label-free imaging and analysis of Pinus pollen grains using optical diffraction tomography," Scientific reports **8**, 1782 (2018).
29. J. Hur, K. Kim, S. Lee, H. Park, and Y. Park, "Melittin-induced alterations in morphology and deformability of human red blood cells using quantitative phase imaging techniques," Scientific reports **7**, 9306 (2017).
30. J. Yoon, Y. Jo, M.-h. Kim, K. Kim, S. Lee, S.-J. Kang, and Y. Park, "Identification of non-activated lymphocytes using three-dimensional refractive index tomography and machine learning," Scientific Reports **7**(2017).
31. T. Tougan, J. R. Edula, E. Takashima, M. Morita, M. Shinohara, A. Shinohara, T. Tsuboi, and T. Horii, "Molecular Camouflage of Plasmodium falciparum Merozoites by Binding of Host Vitronectin to P47 Fragment of SERA5," Scientific reports **8**, 5052 (2018).
32. S. Kwon, Y. Lee, Y. Jung, J. H. Kim, B. Baek, B. Lim, J. Lee, I. Kim, and J. Lee, "Mitochondria-targeting indolizino [3, 2-c] quinolines as novel class of photosensitizers for photodynamic anticancer activity," European journal of medicinal chemistry **148**, 116-127 (2018).
33. K. Kim, S. Lee, J. Yoon, J. Heo, C. Choi, and Y. Park, "Three-dimensional label-free imaging and quantification of lipid droplets in live hepatocytes," Scientific reports **6**, 36815 (2016).
34. J. Lim, K. Lee, K. H. Jin, S. Shin, S. Lee, Y. Park, and J. C. Ye, "Comparative study of iterative reconstruction algorithms for missing cone problems in optical diffraction tomography," Optics express **23**, 16933-16948 (2015).
35. M. Barth, R. K. Bryan, and R. Hegerl, "Approximation of missing-cone data in 3D electron microscopy," Ultramicroscopy **31**, 365-378 (1989).
36. B. Simon, M. Debailleul, M. Houkal, C. Ecoffet, J. Bailleul, J. Lambert, A. Spangenberg, H. Liu, O. Soppera, and O. Haeberlé, "Tomographic diffractive microscopy with isotropic resolution," Optica **4**, 460-463 (2017).
37. M. Lee, S. Shin, and Y. Park, "Reconstructions of refractive index tomograms via a discrete algebraic reconstruction technique," Opt. Express **25**, 27415-27430 (2017).
38. A. Kuś, M. Dudek, B. Kemper, M. Kujawińska, and A. Vollmer, "Tomographic phase microscopy of living three-dimensional cell cultures," in (SPIE, 2014), 8.
39. L. Yu-Chih and C. Chau-Jern, "Sectional imaging of spatially refractive index distribution using coaxial rotation digital holographic microtomography," Journal of Optics **16**, 065401 (2014).
40. S. Vertu, J. Flügge, J.-J. Delaunay, and O. Haeberlé, "Improved and isotropic resolution in tomographic diffractive microscopy combining sample and illumination rotation," Central European Journal of Physics **9**, 969-974 (2011).
41. S. Mojtaba Shakeri, L. J. van Vliet, and S. Stallinga, "Impact of partial coherence on the apparent optical transfer function derived from the response to amplitude edges," Appl. Opt. **56**, 3518-3530 (2017).
42. J. R. Swedlow, J. W. Sedat, and D. A. Agard, "Deconvolution in optical microscopy," in *Deconvolution of images and spectra (2nd ed.)*, A. J. Peter, ed. (Academic Press, Inc., 1996), pp. 284-309.
43. J. Li, Q. Chen, J. Sun, J. Zhang, J. Ding, and C. Zuo, "Three-dimensional tomographic microscopy technique with multi-frequency combination with partially coherent illuminations," Biomedical Optics Express **9**, 2526-2542 (2018).
44. R. Horstmeyer, J. Chung, X. Ou, G. Zheng, and C. Yang, "Diffraction tomography with Fourier ptychography," Optica **3**, 827-835 (2016).
45. J. M. Soto, J. A. Rodrigo, and T. Alieva, "Label-free quantitative 3D tomographic imaging for partially coherent light microscopy," Optics Express **25**, 15699-15712 (2017).
46. A. Descloux, K. S. Grußmayer, E. Bostan, T. Lukes, A. Bouwens, A. Sharipov, S. Geissbuehler, A. L. Mahul-Mellier, H. A. Lashuel, M. Leutenegger, and T. Lasser, "Combined multi-plane phase retrieval and super-resolution optical fluctuation imaging for 4D cell microscopy," Nature Photonics **12**, 165-172 (2018).
47. J. Li, Q. C. J. Sun, J. Zhang, J. Ding, and C. Zuo, "Three-dimensional optical diffraction tomographic microscopy with optimal frequency combination with partially coherent illuminations," arXiv, 1803.01151 (2018).
48. Y. Sung and R. R. Dasari, "Deterministic regularization of three-dimensional optical diffraction tomography," JOSA A **28**, 1554-1561 (2011).
49. S. Lee, K. Lee, S. Shin, and Y. Park, "Generalized image deconvolution by exploiting the transmission matrix of an optical imaging system," Scientific Reports **7**, 8961 (2017).